\documentclass[aps,pre,twocolumn,superscriptaddress]{revtex4-1}

\usepackage{amsmath}
\usepackage{amssymb}
\usepackage{hyperref} 
\usepackage[arrow, matrix, curve]{xy}
\usepackage{bm}
\usepackage{yhmath}
\usepackage{color}

\DeclareMathOperator\Real{Re}
\DeclareMathOperator\Imag{Im}

\usepackage{color}
\newcommand\diff{\mathrm{d}}

\newcommand\mat[1]{\bm{#1}}

\newcommand\norm[1]{\big\|{#1}\big\|}

\usepackage[usenames,dvipsnames]{xcolor}
\hypersetup{colorlinks=true, linkcolor=BrickRed, urlcolor=blue!50!black, citecolor=blue!50!black}

\usepackage[normalem]{ulem}

\makeatletter
\newcommand\hide@visible[1]{%
  \bgroup\fboxsep=.3ex\colorbox{Gray}{begin hide}%
  #1\colorbox{Gray}{end hide}\egroup%
}
\newcommand\hide@hidden[1]{%
  \bgroup\fboxsep=.3ex\colorbox{Gray}{hidden text}%
}
\newcommand\hide@invisible[1]{}
\newcommand\makevisible{\let\hide\hide@visible}
\newcommand\makehidden{\let\hide\hide@hidden}
\newcommand\makeinvisible{\let\hide\hide@invisible}
\makeatother
\makehidden

\begin{document}


\title{ Mode-coupling theory for multiple decay channels}



\author{Simon Lang}
\affiliation{Institut f\"ur Theoretische Physik, Leopold-Franzens-Universit\"at Innsbruck, Technikerstra{\ss}e~25/2,  A-6020 Innsbruck, Austria}
\affiliation{Institut f\"ur Theoretische Physik, Friedrich-Alexander-Universit\"at Erlangen-N\"urnberg, Staudtstra{\ss}e~7, 91058, Erlangen, Germany}

\author{Rolf Schilling}
\affiliation{Institut f\"ur Physik, Johannes Gutenberg-Universit\"at Mainz,
Staudinger Weg 7, 55099 Mainz, Germany}

\author{Thomas Franosch}
\affiliation{Institut f\"ur Theoretische Physik, Leopold-Franzens-Universit\"at Innsbruck, Technikerstra{\ss}e~25/2,  A-6020 Innsbruck, Austria}
\affiliation{Institut f\"ur Theoretische Physik, Friedrich-Alexander-Universit\"at Erlangen-N\"urnberg, Staudtstra{\ss}e~7, 91058, Erlangen, Germany}

\date{\today}



\begin{abstract}
We investigate the properties of a class of  mode-coupling equations for the glass transition where the density mode decays into multiple relaxation  channels. We prove the existence and uniqueness of the solutions for Newtonian as well as Brownian dynamics and demonstrate that they fulfill the requirements of  correlation functions, in the latter case the solutions are purely relaxational.
Furthermore, we  construct an effective mode-coupling functional which allows to map the theory to the case of a single decay channel, such that the covariance principle found for the mode-coupling theory for simple liquids is properly generalized.  This in turn allows  establishing the maximum theorem stating that long-time limits of mode-coupling solutions can be calculated as maximal solutions of a fixed-point equation without relying on the dynamic solutions.
\end{abstract}

\maketitle


\section{Introduction}
The dynamic properties of a liquid are encoded in time-dependent correlation functions of  suitable observables. The principal quantity characterizing the collective motion is the dynamic density-density correlation function or intermediate scattering function $S(q,t)$~\cite{Hansen:Theory_of_Simple_Liquids} which is experimentally accessible either directly by scattering techniques on the molecular scale or can be calculated as statistical average in particle-tracking methods for colloidal systems. Since these correlation functions are subject to some underlying stochastic process, say Newtonian dynamics, probability theory imposes constraints on the class of permissible functions~\cite{Feller:Probability_2}, the most important one being that the corresponding power spectra are non-negative. While these properties are automatically fulfilled in an experiment or a computer simulation in the stationary state or thermal equilibrium, a  theoretical description usually relies on a series of simplifying
assumptions
thereby
necessarily involving  approximations. It is hence a non-trivial question if the solutions of theoretical models respect general features of dynamic correlation functions within the framework of probability theory.

In the case of colloidal particles which are model liquids on the micrometer scale, the microscopic dynamics is Brownian motion rather than Newtonian dynamics. The solvent friction overdamps the dynamics such that all colloidal correlation functions exhibit pure relaxations, i.e. they can be represented as positive superpositions of decaying exponentials only. The corresponding even more restricted class of permissible functions is well-known in the mathematical literature~\cite{Feller:Probability_2}  and the property is referred to as  completely monotone functions. A manifestation of this property is that the power spectra are superpositions of Lorentzians centered at zero frequency.

For dense or supercooled liquids the structural relaxation slows down drastically and a microscopic theory for the plethora of phenomena associated with the glass transition is provided by
 the mode-coupling theory (MCT) of the glass transition for simple liquids developed by G\"otze and collaborators~\cite{Bengtzelius:1984,Goetze:Complex_Dynamics}. The theory
provides a set of integro-differential equations for $S(q,t)$ and  makes a series of non-trivial predictions in the vicinity of a dynamic bifurcation, for example  the emergence of an extended  plateau in the intermediate scattering functions, power-law relaxations at mesoscopic time windows, the time-temperature superposition principle, etc.
Although the sharp dynamical transition predicted by MCT becomes smeared  by ergodicity-restoring processes,  various facets of the theory have been tested successfully~\cite{Goetze:1999,Goetze:Complex_Dynamics} in numerous experiments on supercooled liquids, colloids, and extensive computer simulations
on  hard spheres, Lennard-Jones mixtures, and Yukawa particles.

Many of the  mathematical properties for the intermediate scattering function $S(q,t)$ in the case of simple one-component systems have been worked out rigorously, in particular,
it has been demonstrated  that solutions of the MCT equations uniquely exist both for the Newtonian~\cite{Haussmann:1990} as well as for Brownian dynamics~\cite{Goetze:1995}
compatible with the constraints provided by probability theory. In the Brownian case the solutions are completely monotone thus reflecting the underlying relaxational dynamics~\cite{Goetze:1995}
(see \cite{Franosch:1999} for
explicit numerical results for a distribution of relaxation rates for a hard-sphere system)   and the existence of the long-time limits called glass-form factor follows.
Second,  the MCT equations respect a covariance principle under certain transformations which entails a series of properties for the solutions~\cite{Goetze:Complex_Dynamics}.
For example, the glass-form factor is the maximal solution of a fixed-point equation and it can be determined by employing a convergent monotone iteration scheme and can therefore be calculated without solving the dynamical equations. Glass transition lines are identified with bifurcation singularities for the glass-form factor and the class of possible bifurcations is completely characterized.

The simplest generalization of MCT is to consider multi-component mixtures which requires us to employ matrix-valued correlation functions $S_{\mu\nu}(q,t)$ where the indices $\mu,\nu$ refer to the different species in the liquid. Then the equations of motion couple different species and the theory naturally has to deal with integro-differential equations for matrices. The notion of positivity then has to be adapted to positive-definite matrices and it has been shown that the covariance principle  can be properly generalized~\cite{Franosch:2002} and all properties shown for the single-component system can be proven as well.

Most glass formers are molecular liquids. Rigid molecules display an orientational degree of freedom besides the translational one, and a mode-coupling theory for a molecular liquid~\cite{Scheidsteger:1997,Fabbian:1999} as well as a single molecule in a solvent of spherical particles~\cite{Franosch:1997} has been worked out. Expanding the orientation in a complete set, i.e. spherical harmonics $Y_{\ell m}$ in the case of linear molecules or Wigner rotation matrix $D^{(\ell)}_{mn}$ for arbitrarily-shaped particles again leads to matrix-valued correlation functions, yet a new phenomenon occurs: the density changes both by translation and orientation and to describe the slow structural relaxation, it is necessary to consider multiple relaxation channels. As a non-trivial prediction of these MCT-equations the glassy dynamics is independent of inertial parameters, i.e. the mass and moment of inertia of its molecular constituents~\cite{Franosch:1997,Scheidsteger:1997,Fabbian:1999}.
 However,  the mathematical structure of these MCT equations differs explicitly
from the case of multicomponent systems, in particular the covariance property appears to be lost.

 Recently, liquids confined to a slit have been considered~\cite{Lang:2010,Lang:2012} where translational invariance perpendicular to the walls is explicitly broken. Expansion into suitable Fourier modes for the perpendicular degree of freedom again requires to deal with matrix-valued correlators $S_{\mu\nu}(q,t)$ where the indices now refer to different modes. The fluctuating density decays  by currents perpendicular or parallel to the walls  and the mathematical structure of the MCT is identical to the one of molecules.

The goal of this work is to provide proofs that the MCT equations for multiple relaxation channels display unique solutions which reflect the constraints imposed on being correlation functions. This is achieved in the case of Newtonian as well as Brownian dynamics, in the latter case the theory ignores hydrodynamic interactions. We also formulate a covariance principle for the equations of motion in a wider sense than used for one- or multi-component systems, which is essential to connect to the techniques developed to prove the properties within the mode-coupling approach. We shall show that this generalized covariance principle is suited to introduce an effective mode-coupling functional such that the theory assumes the mathematical structure of a multi-component mixture. The properties of MCT equations for multiple relaxation channels follow from this mapping corroborating that the MCT approach is a robust strategy to arrive at physical correlation functions.

\section{Matrix-valued correlation functions}\label{Sec:correlation_function}

The quantities of interest are the generalized intermediate scattering functions (ISF)
\begin{equation}
S_{\mu\nu}(q,t) = N^{-1} \langle \rho_{\mu}(\vec{q},t)^* \rho_\nu(\vec{q},0) \rangle ,
\end{equation}
as the time-correlation matrix of  a set of density modes $\rho_{\mu}(\vec{q},t)$ labeled by a discrete mode index $\mu$ and a continuous wave number $\vec{q}$. Examples for such density modes will be given below. Furthermore,  $N$ is the number of particles and conventions are such that $S_{\mu\nu}(q,t)$ becomes independent of the system size in the thermodynamic limit.
The theory becomes more elegant by treating the generalized ISF as a  matrix $[\mat{S}(q,t)]_{\mu\nu} = S_{\mu\nu}(q,t)$  and similarly for other correlation functions.
To avoid technical complications we shall assume the matrices to be finite-dimensional and the wave numbers to be of a discrete finite set.  We will suppress the wave number $q$ in the following if it merely serves as label, the relations derived then hold for each wave number separately.

General properties of the ISF can be inferred from the fact that for any set of complex numbers $y_\mu$ (and for each fixed wave number $q$) the quantity $\sum_{\mu\nu} y_\mu^* S_{\mu\nu}(t) y_\nu$ is an autocorrelation function of a stationary stochastic process.
By  the spectral representation theorem~\cite{Feller:Probability_2} it can be written as the characteristic function $\int e^{-\text{i} \Omega t} {R}(\diff \Omega)$ of a finite spectral measure ${R}(\Omega)$ corresponding to a non-decreasing, right-continuous function on the real line with ${R}(-\infty)=0, {R}(\infty)  <\infty$.
Within equilibrium statistical physics the time-reversed stochastic process obeys the same probabilistic law implying that the spectral measure is also symmetric.

The  generalization to the matrix case is straightforward~\cite{Gesztesy:2000}
\begin{equation}\label{eq:spectral_representation}
 \mat{S}(t) = \int e^{-\text{i} \Omega t}  \mat{R}(\diff \Omega),
\end{equation}
such that  $\mat{R}(\Omega)$ is
 a self-adjoint matrix-valued  measure, i.e. $R_{\mu\nu}(\Omega)$ is a complex finite Borel measure on the real line $\Omega\in \mathbb{R}$ and a positive-semidefinite matrix for fixed $\Omega$.
 Thus the matrix correlator  fulfills $\mat{S}(-t) = \mat{S}^\dagger(t)$ and by the equilibrium property it is also invariant under time-reversal $\mat{S}(-t) = \mat{S}(t)$. Combining both properties one infers that the ISF is hermitian $\mat{S}(t) =\mat{S}^\dagger(t)$.

We define the Fourier-Laplace transform by
\begin{equation}
\hat{\mat{S}}(z) = \text{i} \int_0^\infty e^{\text{i} z t} \mat{S}(t) \diff t,
\end{equation}
where the complex frequency $z$  is confined to the upper complex half plane $\mathbb{C}_+ = \{ z \in \mathbb{C} | \text{Im}[z]  > 0 \}$. The spectral representation theorem implies
\begin{equation}
 \hat{\mat{S}}(z) = \int \frac{1 }{\Omega - z}  \mat{R}(\diff \Omega) , \qquad z\in \mathbb{C}_+.
\end{equation}
The following properties follow readily   from the preceding relation for the Fourier-Laplace transforms of matrix correlation functions in equilibrium and complex frequencies $z\in \mathbb{C}_+$
\begin{enumerate}
\item[(1)] $\hat{\mat{S}}(z)$ is analytic
\item[(2)]  $\hat{\mat{S}}( -z^*) = -\hat{\mat{S}}^\dagger(z)$
\item[(3)]  $\lim_{\eta \to \infty} \eta \Imag[\hat{\mat{S}}(z = i\eta)]$ is finite
\item[(4)]  $\Imag[ \hat{\mat{S}}(z)]  \succeq 0$,
\end{enumerate}
where $\Imag{[\hat{\mat{S}}(z)]}=[\hat{\mat{S}}(z)-\hat{\mat{S}}^{\dagger}(z)]/2\text{i}$ is the proper generalization of the imaginary part for  the matrix $\hat{\mat{S}}(z)$,  and the symbol $\succeq 0$ indicates that the matrix is positive-semidefinite.
Conversely,  the Riesz-Herglotz representation theorem~\cite{Gesztesy:2000} reveals that these four conditions  are sufficient for $\hat{\mat{S}}(z)$ being the Laplace transform of an equilibrium matrix correlation function.

An important subclass of equilibrium correlation functions arises if the dynamics is purely relaxational, for example in the case of Brownian dynamics. Then all time-correlations consist of  superpositions of relaxing exponentials in the following sense
\begin{equation}\label{eq:completely_monotone}
 \mat{S}(t) = \int_0^\infty e^{-\gamma t}  \mat{a}(\diff \gamma), \qquad t\geq 0,
\end{equation}
where $\mat{a}(\gamma)$ is again a self-adjoint matrix-valued measure.  Such functions $\mat{S}(t)$ are referred to as completely monotone and display the property that time-derivatives $ [-\partial_t ]^\ell \mat{S}(t) \succeq 0, \ell \in \mathbb{N}$ are positive-semidefinite for $t>0$. By Bernstein's theorem~\cite{Feller:Probability_2} this condition is equivalent to the integral representation, Eq.~\eqref{eq:completely_monotone}.

The Fourier-Laplace transform of a completely monotone matrix-valued function can be extended  to  complex frequencies not located on the negative imaginary axis and is  represented as
\begin{equation}
 \hat{\mat{S}}(z) = \int_0^\infty \frac{-1}{z + i \gamma} \mat{a}(\diff \gamma), \qquad z\in \mathbb{C}\setminus i \mathbb{R}^-.
\end{equation}
The following properties can be verified easily for  $z\in \mathbb{C}\setminus i\mathbb{R}^-$
\begin{enumerate}
\item[$(1^*)$] $\hat{\mat{S}}(z)$ is analytic
\item[$(2^*)$]  $\hat{\mat{S}}( -z^*) = -\hat{\mat{S}}^\dagger(z)$.
\item[$(3^*)$]  $\lim_{\eta \to \infty} \hat{\mat{S}}(z = i\eta) = 0$.
\item[$(4^*)$]  $\Real[ \hat{\mat{S}}(z)] \succeq 0$ for $\Real[z] < 0$,
\end{enumerate}
where  $\Real{[\hat{\mat{S}}(z)]}=[\hat{\mat{S}}(z)+\hat{\mat{S}}^{\dagger}(z)]/2$ now denotes the hermitian part of the matrix $\hat{\mat{S}}(z)$.
Conversely, these four properties guarantee a representation of the form of Eq.~\eqref{eq:completely_monotone} (see Ref.~\cite{Grippenberg:Volterra_Integral}, Sec. 5, Theorem 2.6).

\section{Mode-coupling equations with multiple decay channels}
We consider the dynamics of a generalized fluctuating density $\rho_\mu(\vec{q},t)$
 where $\vec{q}$ is a wave number along a direction where translational invariance in the statistical sense holds. The additional mode index $\mu$ selects further properties of the density mode. For example, in the case of multicomponent mixtures it refers to the different species, whereas for linear molecules it represents a multi-index $\mu = (\ell, m)$ characterizing the orientation of the molecule in terms of spherical harmonics $Y_{\ell m}(\vartheta,\varphi)$. For molecules of arbitrary shape Wigner rotation matrices $D^{(\ell)}_{mn}(\vartheta,\varphi,\chi)$ have to be employed and correspondingly $\mu =(\ell, m, n)$ encodes the dependence on the 3 Euler angles $(\vartheta,\varphi,\chi)$ of the orientation of the molecule. In liquids confined to a slab of effective width $L$ the wave number $\vec{q}$ is only for directions parallel to the walls and $\mu\in \mathbb{Z}$ indicates the discrete Fourier modes of wave numbers $Q_\mu = 2\pi \mu/L$ for the modulation perpendicular
to the walls.

The density modes fulfill a continuity equation
\begin{equation} \label{eq:continuity}
\dot{\rho}_\mu (\vec{q}, t) = i \sum_{\alpha=1}^r \, q^\alpha_\mu \, j^\alpha_\mu (\vec{q}, t),
\end{equation}
where we allow for $r\in \mathbb{N}$ different decay channels. For one- or multi-component bulk liquids comprised of structureless particles only a single current is needed to be considered, $r=1$, and the corresponding currents $j^\alpha_\mu(\vec{q},t)$ are identified with the species-dependent longitudinal mass transport, whereas the real-valued  channel coupling $q^\alpha_\mu$ corresponds to the magnitude of the wave number $q=|\vec{q}|$. For linear molecules already 2 decay channels become relevant~\cite{Franosch:1997,Scheidsteger:1997}, a translational one $\alpha = \text{T}$   with $q^\text{T}_{\ell m} = q$ and an orientational one $\alpha=\text{R}$ where $q^\text{R}_{\ell m} = \sqrt{\ell (\ell +1)}$. General molecules require even 6 decay channels to be considered~\cite{Schilling:2002}, 3 translational ones corresponding to the  cartesian directions  and 3 orientational ones associated with the  cartesian components of the body-fixed frame.
In case of liquids in slit geometry~\cite{Lang:2010,Lang:2012} the decay of the density mode occurs parallel and perpendicular to the walls and characteristic couplings are  $q_\mu^{||} = q$, and $q^\perp_\mu=Q_\mu$, respectively.

Correlating the density modes
$S_{\mu\nu}(q,t)  = N^{-1} \langle \rho_\mu(\vec{q},t)^* \rho_\nu(\vec{q},0) \rangle$ yields matrix-valued intermediate scattering functions which should reflect the properties elaborated in Sec.~\ref{Sec:correlation_function}.
For \emph{Newtonian} dynamics, a formally exact equation of motion can be
derived relying on the Zwanzig-Mori projection operator formalism~\cite{Hansen:Theory_of_Simple_Liquids,Goetze:Complex_Dynamics} choosing as distinguished variables
first  the set of fluctuating densities $\{\rho_\mu (\vec{q}, t) \}$ and then the currents $\{j^\alpha_\mu (\vec{q}, t)\}$. The first equation of motion is found to
\begin{equation}\label{eq:first_eom}
 \dot{\mat{S}}(t) + \int_0^t \mat{K}(t-t') \mat{S}^{-1} \mat{S}(t') \diff t' = 0,
\end{equation}
where $\mat{S} = \mat{S}(t=0)$  is the initial value of the ISF and is identified as generalized static structure factor. We again suppress the dependence on the wave number $q$ if no confusion can occur and all quantities appearing in the equations are to be read as wave number-dependent.
The explicit expression for the memory kernel $[\mat{K}(q,t)]_{\mu\nu}= K_{\mu\nu}(q,t)$ reveals that it corresponds to a
 time-dependent correlation function of the variables $\dot{\rho}_\mu(\vec{q}), \dot{\rho}_\nu(\vec{q})$ such that the dynamics is driven by a reduced Liouville operator~\cite{Hansen:Theory_of_Simple_Liquids,Goetze:Complex_Dynamics}. By Fourier-Laplace transform one can formally solve Eq.~\eqref{eq:first_eom}
\begin{equation}\label{eq:first_zwanzig}
 \hat{\mat{S}}(z) = - \left[ z \mat{S}^{-1} + \mat{S}^{-1} \hat{\mat{K}}(z) \mat{S}^{-1} \right]^{-1} .
\end{equation}

Employing the continuity equation, Eq.\eqref{eq:continuity},
the memory kernel naturally splits
\begin{equation} \label{eq:split}
{K}_{\mu\nu}(q,t)=\sum_{\alpha=1}^{r} \sum_{\beta=1}^{r}  q_{\mu}^{\alpha} {\mathcal{K}}_{\mu\nu}^{\alpha \beta}(q,t) q_{\nu}^{\beta}.
\end{equation}
We  indicate quantities associated with  mode indices and  channel indices by calligraphic symbols and again
use matrix notation, e.g. $[{\bm{\mathcal{K}}}(q,t)]^{\alpha\beta}_{\mu\nu} = {\mathcal{K}}_{\mu\nu}^{\alpha\beta}(q,t)$.

Employing the second  Zwanzig-Mori projection step for the case of  \emph{Newtonian} dynamics one finds the equation of motion for the currents
\begin{equation}\label{eq:second_eom}
 \dot{\bm{\mathcal{K}}}(t) + \bm{\mathcal{J}} \bm{\mathcal{D}}^{-1}  \bm{\mathcal{K}}(t)+  \int_0^t \bm{\mathcal{J}} \bm{\mathcal{M}}(t-t') \bm{\mathcal{K}}(t') \diff t' =0,
\end{equation}
where $\mathcal{J}^{\alpha\beta}_{\mu\nu}(q) = N^{-1} \langle j^\alpha_\mu (\vec{q})^*j^{\beta}_{\nu}(\vec{q}) \rangle$ is the static current correlator matrix which also serves as initial  condition $\bm{\mathcal{K}}(q,t=0) =\bm{\mathcal{J}}(q)$, and $[{\bm{\mathcal{M}}}(q,t)]^{\alpha\beta}_{\mu\nu}=\mathcal{M}^{\alpha \beta}_{\mu\nu} (q,t)$  is a force kernel.
Furthermore, we allow for  an \emph{instantaneous} damping term with  positive-semidefinite matrices $\bm{\mathcal{D}}^{-1}(q) \succeq 0$.

A formal solution is again obtained by Fourier-Laplace transform
\begin{equation}\label{eq:newtonian_dynamics}
 \hat{\bm{\mathcal{K}}}(z) = - \left[ z \bm{\mathcal{J}}^{-1} + i \bm{\mathcal{D}}^{-1} + \hat{\bm{\mathcal{M}}}(z) \right]^{-1}.
\end{equation}
Provided  $\hat{\bm{\mathcal{M}}}(z) = \mathcal{O}(z^{-1})$ for $z\to \infty$,
the leading terms in the high-frequency expansion for the current correlator and the ISF  do not involve the force kernel
\begin{align}\label{eq:high_frequency_newton}
\hat{\mat{K}}(z) =& - z^{-1} \mat{J} + z^{-2}  i \bm{\nu} + \mathcal{O}(z^{-3}), \\
 \hat{\mat{S}}(z) =& -z^{-1} \mat{S} - z^{-3} \mat{J} + i z^{-4} \bm{\nu} + \mathcal{O}(z^{-5}), \quad z\to \infty ,
\end{align}
where the contracted static current correlation matrix is denoted by $J_{\mu\nu}(q) = \sum_{\alpha\beta} q_{\mu}^{\alpha} \mathcal{J}_{\mu\nu}^{\alpha \beta}(q) q_{\nu}^{\beta}$
and similarly $[\bm{\nu}(q)]_{\mu\nu} = \sum_{\alpha\beta} q_{\mu}^{\alpha} [\bm{\mathcal{J}}(q) \bm{\mathcal{D}}^{-1}(q) \bm{\mathcal{J}}(q)]_{\mu\nu}^{\alpha \beta}(q) q_{\nu}^{\beta}$. The corresponding  short-time expansion is readily inferred to
\begin{equation}\label{eq:short_time_Newton}
{\mat{S}}(t) =  \mat{S} -  \mat{J} t^2/2 +  \bm{\nu} t^3/6 +  \mathcal{O}(t^{4}), \qquad t\to 0.
\end{equation}
Here
the leading correction of order ${\cal O}(t^4)$ depend on the force kernel $\bm{\mathcal{M}}(t=0)$.

Within the  MCT approach the memory kernel ${\bm{\mathcal{M}}}(q,t)$ is approximated
as a  functional that is local in time of the ISF
as positive superpositions of products of the ISF~\cite{Lang:2012}
\begin{align} \label{eq:MCT_M}
\mathcal{M}_{\mu\nu}^{\alpha \beta}(q,t) = & {\cal F}^{\alpha\beta}_{\mu\nu}[ \mat{S}(t), \mat{S}(t);q]+(\mathcal{M}_{\text{reg}})^{\alpha\beta}_{\mu \nu}(q,t).
\end{align}

Explicitly the MCT functional is represented as
\begin{align}\label{eq:MCT_Functional}
{\cal F}^{\alpha\beta}_{\mu\nu}[ \bm{S}(t), \bm{S}(t);q]
=& \frac{1}{2N} \sum_{\vec{q}_{1},\vec{q}_{2}=\vec{q}-\vec{q}_{1}} \sum_{\mu_{1}\mu_{2} \atop \nu_{1} \nu_{2}} \mathcal{Y}^\alpha_{\mu,\mu_{1} \mu_{2}}(\vec{q},\vec{q}_{1}\vec{q}_{2}) \nonumber \\
&\times S_{\mu_{1}\nu_{1}}(q_{1},t) S_{\mu_{2}\nu_{2}}(q_{2},t) \mathcal{Y}^\beta_{\nu,\nu_{1} \nu_{2}}(\vec{q},\vec{q}_{1}\vec{q}_{2})^*,
\end{align}
and the  vertices $\mathcal{Y}^\alpha_{\mu, \mu_1\mu_2}(\vec{q},\vec{q}_1\vec{q}_2)$ are determined by static properties of the liquid only and assumed to be known smooth functions of control parameters, see Ref.~\cite{Lang:2012}. The force correlator may contain an additional regular damping kernel $(\bm{\mathcal{M}}_{\text{reg}})(q,t)$ encoding faster dynamical processes, which are not captured by the MCT-functional. This kernel is assumed to be known a priori and satisfies the constraints formulated in Sec.~\ref{Sec:correlation_function} of an equilibrium matrix-valued correlation function.

Equations (\ref{eq:first_eom}),(\ref{eq:split}),(\ref{eq:second_eom}), and (\ref{eq:MCT_M}),(\ref{eq:MCT_Functional}) together with the initial conditions $\mat{S}(t=0) = \mat{S}, \bm{\mathcal{K}}(t=0) = \bm{\mathcal{J}}$ constitute a closed set of equations for the Newtonian case, and we shall prove the existence and uniqueness of solutions that are correlation functions in the next section.

For \emph{Brownian} dynamics Eq.~\eqref{eq:newtonian_dynamics} is replaced by
\begin{equation}\label{eq:Brownian_dynamics}
 \hat{\bm{\mathcal{K}}}(z) = - \left[ i \bm{\mathcal{D}}^{-1} + \hat{\bm{\mathcal{M}}}(z) \right]^{-1},
\end{equation}
such that $\bm{\mathcal{D}}$ is a positive-definite matrix. Here again the memory-kernel may contain as well an a priori known regular damping, which satisfies the additional constraints of a purely relaxational equilibrium matrix-valued correlation function, see Sec.~\ref{Sec:correlation_function}. By the same arguments as above the high-frequency expansion for the ISF is found to
\begin{equation}
 \hat{\mat{S}}(z) = -z^{-1} \mat{S} + z^{-2} i \mat{D} + {\cal O}(z^{-3}), \qquad z\to \infty ,
\end{equation}
with the contracted relaxation rate matrix $D_{\mu\nu}(q) = \sum_{\alpha\beta} q_{\mu}^{\alpha} \mathcal{D}_{\mu\nu}^{\alpha \beta}(q) q_{\nu}^{\beta}$. In the temporal domain this implies
\begin{equation}
{\mat{S}}(t) =  \mat{S} -  \mat{D} t + {\cal O}(t^{2}), \qquad t\to 0 ,
\end{equation}
and the positive definite matrix $\mat{D}$ determines the initial decay of the ISF.

From Eq.~\eqref{eq:Brownian_dynamics} one infers   $\hat{\bm{\mathcal{K}}}(z) \to  i \bm{\mathcal{D}}$ as $z\to \infty$ which suggests, that $\bm{\mathcal{K}}(t)$
contains a $\delta$-function at the time origin. To arrive at integro-differential equations with smooth functions, we split off the high-frequency limit,
$ \delta \hat{\bm{\mathcal{K}}}(z) = \hat{\bm{\mathcal{K}}}(z) - i \bm{\mathcal{D}}$, consistent with  its contraction $\delta \hat{\mat{K}}(z) =\hat{\mat{K}}(z) - i \mat{D}$. Then
Eq.~\eqref{eq:first_zwanzig} translates back in the temporal domain to
\begin{equation}\label{eq:first_eom_Brown}
 \dot{\mat{S}}(t) + \mat{D} \mat{S}^{-1} \mat{S}(t) + \int_0^t \delta \mat{K}(t-t') \mat{S}^{-1} \mat{S}(t') \diff t' = 0.
\end{equation}
Second, Eq.~\eqref{eq:Brownian_dynamics}  implies
\begin{equation}
\left[
 i \bm{\mathcal{D}}^{-1} + \hat{\bm{\mathcal{M}}}(z) \right] \delta\hat{\bm{\mathcal{K}}}(z) = - \hat{\bm{\mathcal{M}}}(z) i \bm{\mathcal{D}}   ,
\end{equation}
which yields an integral equation in the temporal domain
\begin{equation}\label{eq:second_eom_Brown}
  \delta \bm{\mathcal{K}}(t) + \int_0^t \bm{\mathcal{D}} \bm{\mathcal{M}}(t-t') \delta \bm{\mathcal{K}}(t') \diff t' = - \bm{\mathcal{D}}\bm{\mathcal{M}}(t) \bm{\mathcal{D}}.
\end{equation}
Equations (\ref{eq:first_eom_Brown}),(\ref{eq:split}),(\ref{eq:second_eom_Brown}),  and (\ref{eq:MCT_M}),(\ref{eq:MCT_Functional})  together with the initial conditions $\mat{S}(t=0) = \mat{S}$ constitute
a closed set of equations for the Brownian case. The properties of their solutions is one of the central questions of this work.

\section{Construction of MCT solutions}
The existence of MCT solutions with the required properties is intimately connected to the representation of ISF in terms of continued fractions as prescribed by the Zwanzig-Mori approach as well as with the closure via a microscopically derived MCT functional. We shall show that the force kernel obtained via a MCT functional of correlation functions and   completely monotone functions has the same properties as correlation functions and completely monotone function, respectively, and given a force kernel with these properties the ISF as obtained via the continued fraction has again  the same properties.
This observation will allow us to construct an iteration scheme not leaving the class of correlation functions or completely monotone functions. Since both classes are closed the limiting functions, provided they exist, still display the desired properties.
The scheme is then translated to the time domain and shown to be a constructive proof of a unique solution.

\subsection{Properties of the MCT functional}\label{sec:functional}
In the first step we collect some properties of the MCT functional, some of which have been already discussed in Ref.~\cite{Lang:2012}. To avoid technical problems we consider all matrices to be finite-dimensional and the set of wave numbers as a discrete finite set. Then the MCT functional  maps a set of  matrices in the mode space labeled by a wave number to matrices with a mode and channel indices with the same set of labels for the wave numbers. Since the functional is local in time, time merely enters as a parameter.

It is  convenient to introduce the pair-mode indices $a:=(\mu_{1},\mu_{2})$ and $b:=(\nu_{1},\nu_{2})$
and the double indices $\gamma := (\alpha,\mu), \delta:= (\beta,\nu)$. The MCT functional, Eq.~\eqref{eq:MCT_Functional}, can then be represented as
\begin{align}\label{eq:superhyper}
 &\mathcal{F}[\mat{F},\mat{E};q]^{\gamma\delta}=\frac{1}{4N} \sum_{ab} \sum_{\vec{q}_{1},\vec{q}_{2}=\vec{q}-\vec{q}_{1}}\mathcal{Y}^{\gamma}_{ a}(\vec{q},\vec{q}_{1}\vec{q}_{2})\nonumber\\
& \times [\mat{F}(q_{1})\otimes\mat{E}(q_{2})+\mat{E}(q_{1})\otimes\mat{F}(q_{2}) ]_{ab}\mathcal{Y}^{\delta}_{ b}(\vec{q},\vec{q}_{1}\vec{q}_{2})^*.
\end{align}
where $\otimes$ denotes the Kronecker product  in the space of  mode indices, $E_{\mu_{1}\nu_{1}}(q_{1})F_{\mu_{2}\nu_{2}}(q_{2})=[\mat{E}(q_{1})\otimes \mat{F}(q_{2})]_{a=(\mu_{1},\mu_{2}),b=(\nu_{1},\nu_{2})}$.

Positive definiteness of a hermitian matrix $\mat{E}$ shall be indicated $\mat{E} \succ 0$, and positive semi-definiteness by $\mat{E} \succeq 0$.  Similarly for two hermitian matrices we write $\mat{F} \succ \mat{E}$ if $\mat{F}-\mat{E} \succ 0$, etc.
In the following the label $q$ for the wave number is suppressed and all operations are understood component-wise for each $q$. Then one easily shows~\cite{Lang:2012} that $\mat{E} \succeq 0, \mat{F} \succeq 0$ implies $\mat{\mathcal{F}}[\mat{F},\mat{E}] \succeq 0$. Furthermore the functional preserves ordering: $\mat{F} \succeq \mat{E}$ implies $\mat{\mathcal{F}}[\mat{F},\mat{F}] \succeq \mat{\mathcal{F}}[\mat{E},\mat{E}]$.

A matrix-valued correlation function $S_{\mu\nu}(q,t)$ fulfills  $\sum_{i,j} \sum_{\mu\nu}\xi_{i\mu}^*  S_{\mu\nu}(q,t_i-t_j)  \xi_{j\nu} \geq 0$  for any finite set of times $t_i\in \mathbb{R}$ and complex numbers $\xi_{i\mu} \in \mathbb{C}$. By Bochner's theorem~\cite{Feller:Probability_2} this is an equivalent characterization of a  (matrix-valued) correlation function. Then one concludes that if $S_{\mu\nu}(q,t)$ and $T_{\mu\nu}(q,t)$ are matrix-valued correlation functions then so is the matrix $[\mat{S}(q_1,t) \otimes \mat{T}(q_2,t)]_{ab}$.  This follows literally as in the case for scalars by decorating the complex numbers by indices.  For
complex numbers  $s^\gamma_i$  then
\begin{align}\label{eq:MCT_correlation}
& \sum_{i,j} \sum_{\gamma\delta}  s^\gamma_i {}^* \mathcal{F}[\mat{S}(t_i-t_j),\mat{T}(t_i-t_j);q]^{\gamma\delta} s^\delta_j  = \nonumber \\
&= \frac{1}{4N}  \sum_{\vec{q}_{1},\vec{q}_{2}=\vec{q}-\vec{q}_{1}} \Big\{ \sum_{i,j}  \sum_{ab}  \left( \sum_{\gamma}  s^\gamma_i {}^*\mathcal{Y}^{\gamma}_{ a}(\vec{q},\vec{q}_{1}\vec{q}_{2}) \right) \nonumber\\
& \times [\mat{S}(q_{1},t_i-t_j)\otimes\mat{T}(q_{2},t_i-t_j)+  \nonumber \\
&\qquad \mat{T}(q_{1},t_i-t_j)\otimes\mat{S}(q_{2},t_i-t_j) ]_{ab} \left(\sum_\delta  s^\delta_j \mathcal{Y}^{\delta}_{ b}(\vec{q},\vec{q}_{1}\vec{q}_{2})^*\right) \Big\} \geq 0, \nonumber \\
\end{align}
since  the sums in the curly bracket are already non-negative by the characterization of correlation functions. Therefore the MCT functional maps correlation functions to correlation functions.

In the case of \emph{Brownian} dynamics the correlation functions $S_{\mu\nu}(q,t)$ are completely monotone and by Bernstein's theorem~\cite{Feller:Probability_2} can be uniformly approximated for $t\geq 0$ by sums
\begin{equation}
\sum_i e^{-\gamma_i t} \Delta a_{\mu\nu}^{(q)}(\gamma_i) \to  S_{\mu\nu}(q,t),
\end{equation}
with $\Delta \mat{a}^{(q)}(\gamma_i) \succeq 0$ and $\gamma_i \geq 0$. Using such approximants for $\mat{S}(q,t)$ and $\mat{T}(q,t)$ a calculation similar to Eq.~\eqref{eq:MCT_correlation}
shows that the  approximants to $ \mathcal{F}[\mat{S}(t),\mat{T}(t);q]^{\gamma\delta}$ are of the same type. Passing to the limits shows that the MCT functional maps completely monotone correlation functions on completely monotone correlation functions.

\subsection{Zwanzig-Mori's equation of motion}
Next we show that the representation of   the ISF  via memory kernels implies that correlation functions map to correlation functions and in the Brownian case the property of complete monotonicity is also inherited. The proof is performed most easily in the Fourier-Laplace domain and follows essentially  the argument for mixtures~\cite{Franosch:2002}. Since the wave number merely serves as a label, we shall omit the dependence on $q$ in the remainder of this section when convenient.

In the \emph{Newtonian} case we shall  assume that $\bm{\mathcal{M}}(t)$ is a correlation functions which is equivalent to  $\hat{\bm{\mathcal{M}}}(z)$ fulfilling the properties (1)-(4) of Sec.~\ref{Sec:correlation_function}.
The spectral representation shows that $\hat{\bm{\mathcal{M}}}(z= i \eta) = {\cal O}(\eta^{-1})$ as $\eta\to \infty$.
From the second equation of motion, Eq.~\eqref{eq:newtonian_dynamics}, one infers readily that $\hat{\bm{\mathcal{K}}}(z)$ inherits property (1), since the denominator has positive imaginary part (in the matrix sense) for $z\in \mathbb{C}_+$ and hence no zeros can occur. Second, property (2) is readily checked, and by $\hat{\bm{\mathcal{K}}}(z=i\eta) = i \eta^{-1} \bm{\mathcal{J}}
+ {\cal O}(\eta^{-2})$ for $\eta\to \infty$ property (3) follows.
Decomposing the current correlator into real and imaginary parts with  $\Imag[z] > 0$ yields
\begin{align}\label{eq:trick}
& \Imag[\hat{\bm{\mathcal{K}}}(z)] = \hat{\bm{\mathcal{K}}}^\dagger(z) \Imag[-\hat{\bm{\mathcal{K}}}(z)^{-1} ] \hat{\bm{\mathcal{K}}}(z) \nonumber \\
& = \hat{\bm{\mathcal{K}}}^\dagger(z) \left\{ \Imag[z] \bm{\mathcal{J}}^{-1} +  \bm{\mathcal{D}}^{-1} +
 \Imag[\hat{\bm{\mathcal{M}}}(z)]\right\}
\hat{\bm{\mathcal{K}}}(z) \succeq 0 ,
\end{align}
which shows that also property (4) is fulfilled. Hence $\hat{\bm{\mathcal{K}}}(z)$ is again the Fourier-Laplace transform of a correlation function and so is its contraction $\hat{\mat{K}}(z)$, in particular  $\Imag[\mat{K}(z)] \succeq 0$ for $\Imag[z] > 0$. The first equation of motion, Eq.~\eqref{eq:first_zwanzig} shows by the same arguments as above that $\hat{\mat{S}}(z)$ again fulfills properties (1),(2), and (3). The same calculation as above
\begin{align}
&  \Imag[\hat{\mat{S}}(z)] =  \nonumber \\
& \hat{\mat{S}}^\dagger(z) \left\{ \Imag[z] \mat{S}^{-1} + \mat{S}^{-1} \Imag[\hat{\mat{K}}(z) ] \mat{S}^{-1} \right\} \hat{\mat{S}}(z),
\end{align}
demonstrates property (4), $\Imag[\hat{\mat{S}}(z)] \succeq 0$ for $\Imag[z]> 0$. Hence the Zwanzig-Mori equations for the Newtonian case yield correlation functions provided the force kernel is itself a correlation function. For a converse representation theorem see App.~\ref{appendix:representation}.

In the case of \emph{Brownian} dynamics, $\hat{\bm{\mathcal{M}}}(z)$ is assumed to fulfill $(1^*)$-$(4^*)$. Then
Eq.~\eqref{eq:Brownian_dynamics} shows that $- \delta \hat{\bm{\mathcal{K}}}(z) = i \bm{\mathcal{D}}-  \hat{\bm{\mathcal{K}}}(z)$ also satisfies $(1^*)$, $(2^*)$, and $(3^*)$. With the same calculation as in Eq.~\eqref{eq:trick} one can show
\begin{equation}
 \Real[ -\delta\hat{\bm{\mathcal{K}}}(z)] = \hat{\bm{\mathcal{K}}}^\dagger(z) \Real[\hat{\bm{\mathcal{M}}}(z)] \hat{\bm{\mathcal{K}}}(z) .
\end{equation}
Hence for $\Real[z]< 0$, $\Real[-\delta\hat{\bm{\mathcal{K}}}(z)]\succeq  0$ completing the argument that $-\delta\hat{\bm{\mathcal{K}}}(z)$ is completely monotone and so is its contraction $-\delta \hat{\mat{K}}(z)$.
The first Zwanzig-Mori equation, Eq.~\eqref{eq:first_zwanzig}, shows that $\hat{\mat{S}}(z)$ also inherits properties $(1^*)$, $(2^*)$, and $(3^*)$. Using $\Real[\hat{\mat{K}}(z)] =  \Real[\delta\hat{\mat{K}}(z)]$, the relation
\begin{align}
 &\Real[\hat{\bm{S}}(z)] = \nonumber \\
& \hat{\mat{S}}^\dagger(z) \left\{ -\Real[z] \mat{S}^{-1} + \mat{S}^{-1} \Real[-\delta\hat{\mat{K}}(z)] \mat{S}^{-1} \right\}  \hat{\mat{S}}(z),
\end{align}
 then proves also property $(4^*)$. Thus the Zwanzig-Mori equations for Brownian dynamics yield completely monotone correlation functions provided the force kernel is of the same type.

\subsection{Iteration scheme}
The fact that  both the mapping from (completely monotone) correlation functions yields a memory kernel by mode-coupling functional that has the same properties, as well as the mapping from the memory kernel to the ISF again preserving the properties, suggests to construct an iteration scheme within the closed space of such functions.
Such a scheme was first used  by G\"otze and L\"ucke~\cite{Goetze:1976} for the dynamic structure factor of liquid helium.
All operations are again to be understood for each wave number $q$.

In the \emph{Newtonian} case the sequence is initialized by $\left( \mat{S}^{(0)}(t), \bm{\mathcal{K}}^{(0)}(t) \right) = \left( \mat{S}, \bm{\mathcal{J}} \right)$ which are trivially correlation functions. Equivalently in the Fourier-Laplace domain $\left( \hat{\mat{S}}^{(0)}(z), \hat{\bm{\mathcal{K}}}^{(0)}(z) \right) = \left( -\mat{S}/z, -\bm{\mathcal{J}}/z \right)$.
Then the following mapping is employed
\begin{align}
 \hat{\mat{S}}^{(n+1)}(z) =&  - \left[ z \mat{S}^{-1} + \mat{S}^{-1} \hat{\mat{K}}^{(n)}(z) \mat{S}^{-1} \right]^{-1} \label{eq:ISF_iteration}, \\
 \hat{\bm{\mathcal{K}}}^{(n+1)}(z) =& - \left[ z \bm{\mathcal{J}}^{-1} + i \bm{\mathcal{D}}^{-1} +
\hat{\bm{\mathcal{M}}}^{(n)}(z) \right]^{-1} . \label{eq:Newtonian_iteration}
\end{align}
Here $\hat{\mat{K}}^{(n)}(z)$ is of course the contraction of $\hat{\bm{\mathcal{K}}}^{(n)}(z)$ and $\hat{\bm{\mathcal{M}}}^{(n)}(z)$ is the Fourier-Laplace transform of
\begin{align}
 \bm{\mathcal{M}}^{(n)}(q,t) = \bm{\mathcal{F}}[ \mat{S}^{(n)}(t), \mat{S}^{(n)}(t);q]+(\bm{\mathcal{M}}_{\text{reg}})(q,t) .
\end{align}
Hence the coupling of different wave numbers emerges via the MCT functional only.

In the case of \emph{Brownian} dynamics the sequence is started by $\left( \hat{\mat{S}}^{(0)}(z), \hat{\bm{\mathcal{K}}}^{(0)}(z) \right) = \left( -\mat{S}/z, i \bm{\mathcal{D}} \right)$
and Eq.~\eqref{eq:Newtonian_iteration} is replaced by
\begin{align}
  \hat{\bm{\mathcal{K}}}^{(n+1)}(z) =& - \left[ i \bm{\mathcal{D}}^{-1} + \hat{\bm{\mathcal{M}}}^{(n)}(z) \right]^{-1}. \label{eq:Brownian_iteration}
\end{align}
In the preceding subsection it was shown that this corresponds to  $-\delta \hat{\bm{\mathcal{K}}}^{(n+1)}(z) = i \bm{\mathcal{D}} - \hat{\bm{\mathcal{K}}}^{(n+1)}(z)$ which is the Fourier-Laplace transform of a  completely monotone correlation function.

Provided the sequence in the temporal domain $\left( {\mat{S}}^{(n)}(t), {\bm{\mathcal{K}}}^{(n)}(t) \right)$ converges uniformly on each finite time interval, the limits
${\mat{S}}(t) = \lim_{n\to \infty} {\mat{S}}^{(n)}(t)$ are matrix-valued correlation functions which are also completely monotone in the case of Brownian dynamics~\cite{Feller:Probability_2}. Furthermore the limit
represents a solution of the MCT equations.

\subsection{Proof of convergence}
The sequence constructed in the last subsection will be translated to a sequence of functions in the temporal domain.  The convergence will then be demonstrated by providing estimates with respect to a suitable norm that we introduce below following the strategy of Haussmann~\cite{Haussmann:1990}. The main difference to the Picard iteration is that existence and uniqueness can be proven globally with uniform convergence on finite time intervals. The required ingredient is precisely the property that all approximants are correlation functions and therefore bounded by their respective initial values.

First for \emph{Newtonian} dynamics, Eqs.~\eqref{eq:ISF_iteration},~\eqref{eq:Newtonian_iteration}, can be recast to integro-differential equations, similar to Eqs.~\eqref{eq:first_eom},~\eqref{eq:second_eom}. Integrating once,
\begin{align}
 \mat{S}^{(n+1)}(t) =& \mat{S} - \int_0^t\!\!\!\diff t' \!\!\int_0^{t'}\!\!\!\diff t'' \mat{K}^{(n)}(t'-t'') \mat{S}^{-1} \mat{S}^{(n+1)}(t'') ,\label{eq:Picard1}\\
\bm{\mathcal{K}}^{(n+1)}(t) =& \bm{\mathcal{J}} - \int_0^t\!\!\diff t' \bm{\mathcal{J}} \bm{\mathcal{D}}^{-1} \bm{\mathcal{K}}^{(n+1)}(t') \nonumber \\
&  - \int_0^t\!\!\!\diff t' \!\!\int_0^{t'}\!\!\!\diff t'' \bm{\mathcal{J}} \bm{\mathcal{M}}^{(n)}(t'-t'') \bm{\mathcal{K}}^{(n+1)}(t'') , \label{eq:Picard2}
\end{align}
we obtain equations which can be analyzed by the techniques familiar from the Picard iteration scheme for ordinary differential equations. In contrast to this technique, at each iteration step a self-consistent integral equation requires to be solved to evaluate the subsequent approximant. This procedure ensures that each approximant satisfies the constraints of being correlation functions, as it is formulated in Sec.~\ref{Sec:correlation_function}.

Following Ref.~\cite{Franosch:2002}, we denote   the  space  of matrices with mode indices   by $\mathcal{A}_0$ equipped with the standard operator norm $\norm{\cdot}$. Then $\mathcal{A}_0$ is a $C^*$ algebra
with respect to matrix multiplication and hermitian conjugation as $*$-operation. For a finite set of $M$ wave numbers, the matrices naturally form vectors $\mat{E} = [\mat{E}(q) ]_{q=1,\ldots,M} \in \mathcal{A}_0^M$.  Equipping $\mathcal{A}_0^M$ with the maximum norm $\norm{\mat{E}}  = \text{max}_q \norm{ \mat{E}_q }$ and taking all operations component-wise, one easily shows that $\mathcal{A}_0^M$ is again a $C^*$ algebra~\cite{Conway:Functional_Analysis}. Elements are called positive-semidefinite $\mat{E} \succeq 0$ if all components $\mat{E}(q) \succeq 0$ are positive-semidefinite for each $q$. In particular, the norm preserves ordering, $\mat{E} \succeq \mat{F} $ implies $\norm{\mat{E}} \geq \norm{\mat{F} }$.
The space of $M-$ tuples  of all matrices with both mode and channel indices denoted by $\mathcal{A}_1^M$ will be equipped by another $C^*$ algebra by the same procedure and for simplicity we use the same symbol to indicate norms. Then contraction with the selectors induces a Lipschitz-continuous mapping between $\mathcal{A}_1^M \to \mathcal{A}_0^M$.

Since all approximants $\mat{S}^{(n)}(t), \bm{\mathcal{K}}^{(n)}(t), \mat{K}^{(n)}(t)$ are correlation functions, the spectral representation, Eq.~\eqref{eq:spectral_representation},  reveals that
they can not exceed their initial value, i.e. $ \mat{S} \succeq \mat{S}^{(n)}(t)$, $ \bm{\mathcal{J}} \succeq \bm{\mathcal{K}}^{(n)}(t), \mat{J} \succeq \mat{K}^{(n)}(t)$,
which implies $\norm{ \mat{S}^{(n)}(t) } \leq \norm{ \mat{S} }, \norm{\bm{\mathcal{K}}^{(n)}(t) } \leq \norm{ \bm{\mathcal{J}} }, \norm{ \mat{K}^{(n)}(t) } \leq \norm{ \mat{J} }$.

From Eq.~\eqref{eq:Picard1} one derives the standard estimates for $n\in \mathbb{N}$
\begin{align}
\lefteqn{  \norm{ \mat{S}^{(n+1)}(t) - \mat{S}^{(n)}(t) } \leq \nonumber } \\
\leq & \int_0^t \! \diff t'\! \int_0^{t'}\! \diff t'' \norm{ \mat{J} } \,  \norm{ \mat{S}^{-1} }  \,  \norm{ \mat{S}^{(n+1)}(t'') - \bm{S}^{(n)}(t'') }  \nonumber \\
&+
\int_0^t \!\diff t' \! \int_0^{t'} \!\diff t'' \norm{ \mat{S} }  \,  \norm{ \mat{S}^{-1} } \,   \norm{ \mat{K}^{(n)}(t'') - \mat{K}^{(n-1)}(t'') } ,
\end{align}
where the symmetry of the convolution has been exploited to shift the time argument in the last line.
Similarly, Eq.~\eqref{eq:Picard2} yields for $n\in \mathbb{N}$
\begin{align}
\lefteqn{ \norm{ \bm{\mathcal{K}}^{(n+1)}(t) - \bm{\mathcal{K}}^{(n)}(t) } \leq \nonumber } \\
\leq &\int_0^t \diff t' \norm{ \bm{\mathcal{J}} } \, \norm{ \bm{\mathcal{D}}^{-1} } \,  \norm{ \bm{\mathcal{K}}^{(n+1)}(t') - \bm{\mathcal{K}}^{(n)}(t') } \nonumber \\
& + \int_0^t \!\diff t' \int_0^{t'}\! \diff t'' \norm{ \bm{\mathcal{J}} }  \, \norm{ \bm{\mathcal{M}} }  \,  \norm{ \bm{\mathcal{K}}^{(n+1)}(t'') - \bm{\mathcal{K}}^{(n)}(t'') }  \nonumber \\
&+
\int_0^t \diff t'\! \int_0^{t'} \! \diff t'' \norm{ \bm{\mathcal{J}} }^2 \,     \norm{ \bm{\mathcal{M}}^{(n)}(t'') - \bm{\mathcal{M}}^{(n-1)}(t'') } ,
\end{align}
where $\bm{\mathcal{M}} = \bm{\mathcal{M}}(t=0)$. Note, that $\norm{\bm{\mathcal{M}}^{(n)}(t)} \leq \norm{\bm{\mathcal{M}}}$ has been used as well as $\norm{\mat{S}^{(n)}(t)} \leq \norm{\mat{S}}$.

By Lipschitz-continuity of the contraction $ \norm{ \mat{K}^{(n+1)}(t) - \mat{K}^{(n)}(t) } \leq L_1 \norm{ \bm{\mathcal{K}}^{(n+1)}(t) - \bm{\mathcal{K}}^{(n)}(t) }$ and Lipschitz-continuity of the mode-coupling functional
$  \norm{ \bm{\mathcal{M}}^{(n+1)}(t) - \bm{\mathcal{M}}^{(n)}(t) }  \leq L_2  \norm{ \mat{S}^{(n+1)}(t) - \mat{S}^{(n)}(t) }$ with suitable Lipschitz constants $L_1>0,L_2>0$.
Then the growth of the scalar $X^{(n)}(t) = \norm{ \mat{S}^{(n+1)}(t) - \mat{S}^{(n)}(t) } + c \norm{ \bm{\mathcal{K}}^{(n+1)}(t) - \bm{\mathcal{K}}^{(n)}(t) }$, where $c>0$ accounts for the different dimensional units, in the finite time interval $0 \leq t \leq T < \infty$ is controlled by
\begin{align}\label{eq:growth_control}
 X^{(n)}(t) \leq L
\int_0^t \diff t' [ X^{(n)}(t') + X^{(n-1)}(t') ] , \qquad n\in\mathbb{N}
\end{align}
with $L = T \norm{ \mat{J} } \, \norm{ \mat{S}^{-1} } +c \norm{ \bm{\mathcal{J}} } \,  \norm{ \bm{\mathcal{D}}^{-1} } + cT \norm{ \bm{\mathcal{J}} } \, \norm{ \bm{\mathcal{M}} }
+ L_1 T \norm{ \mat{S} } \, \norm{  \mat{S}^{-1} } + cL_2 T \norm{  \bm{\mathcal{J}} }^2$.  The preceding equation
with the additional bound $X^{(n)}(t) \leq K := 2c \norm{ \bm{\mathcal{J}} } + 2 \norm{ \mat{S} }$ is essentially the same as studied in Refs.~\cite{Haussmann:1990,Goetze:1995,Goetze:Complex_Dynamics}, and for completeness we repeat the argument. The sequence $X^{(n)}(t)$ is bounded by the
$a^{(n)}(t)$  recursively defined  by $a^{(n)}(t) = 2 L\int_0^t \diff t' a^{(n-1)}(t'), a^{(0)}(t) = K$. First, $X^{(0)}(t) \leq a^{(0)}(t) =K$ and $X^{(1)}(t) \leq a^{(0)}(t)$ and $X^{(1)}(t) \leq a^{(1)}(t) = 2 L K t$ are obvious. Then
$X^{(2)}(t) \leq 2 L \int_0^t \diff t' a^{(0)}(t') =  a^{(1)}(t)$ which in turn implies $X^{(2)}(t) \leq 2 L \int_0^t \diff t' a^{(1)}(t') = a^{(2)}(t)$. The procedure can be continued to show that $X^{(n)}(t) \leq a^{(m)}(t)$ for $m\leq n$. Explicit calculation shows $a^{(n)}(t) = K (2 L t)^n/n!$ such that  $\norm{\mat{S}^{(n)}(t) - \mat{S}^{(m)}(t) } \leq \sum_{k=m}^{n-1}
 \norm{ \mat{S}^{(k+1)}(t) - \mat{S}^{(k)}(t) }
\leq
\sum_{k= m}^\infty X^{(k)}(t) \leq \sum_{k= m}^\infty a^{(k)}(t) \to 0$ as $n>m \to \infty$ implying $\mat{S}^{(n)}(t)$ being Cauchy sequences  uniformly on every finite time intervals. Since $\mathcal{A}_0^M$ is complete the limit $\mat{S}(t) = \lim_{n\to\infty} \mat{S}^{(n)}(t)$ exists and represents again a correlation function~\cite{Feller:Probability_2}, and similarly for the currents $\bm{\mathcal{K}}(t) = \lim_{n\to \infty} \bm{\mathcal{K}}^{(n)}(t)$.

Uniqueness is demonstrated similarly. Assuming two sets of solutions $\mat{S}(t), \bm{\mathcal{K}}(t)$ and $\tilde{\mat{S}}(t), \tilde{\bm{\mathcal{K}}}(t)$ with the same initial conditions, going through the same steps as above reveals that   $Y(t) := \norm{ \mat{S}(t) - \tilde{\mat{S}}(t) } + c \norm{ \bm{\mathcal{K}}(t) - \tilde{\bm{\mathcal{K}}}(t) }$ satisfies $Y(t) \leq  2 L \int_0^t \diff t' Y(t')$ and $Y(0)=0$. Then by the same argument as in the preceding paragraph induction shows $0\leq Y(t) \leq a^{(n)}(t) \to 0$ as $n\to \infty$ for all times $t$, hence the two solutions are equal.

In the case of \emph{Brownian} dynamics, Eq.~\eqref{eq:Picard1},~\eqref{eq:Picard2} will be replaced by (c.f. Eqs.~\eqref{eq:first_eom_Brown} and~\eqref{eq:second_eom_Brown})
\begin{align}
  \mat{S}^{(n+1)}(t) =& \mat{S} - \int_0^t \mat{D} \mat{S}^{-1} \mat{S}^{(n+1)}(t')\diff t' \nonumber \\
&- \int_0^t\!\!\!\diff t' \!\!\int_0^{t'}\!\!\!\diff t'' \delta \mat{K}^{(n)}(t'-t'') \mat{S}^{-1} \mat{S}^{(n+1)}(t'') ,\label{eq:Picard1_Brownian}\\
\delta \bm{\mathcal{K}}^{(n+1)}(t) =& -\bm{\mathcal{D}} \bm{\mathcal{M}}^{(n)}(t) \bm{\mathcal{D}}  \nonumber \\
&  - \int_0^t\!\!\!\diff t'  \bm{\mathcal{D}} \bm{\mathcal{M}}^{(n)}(t-t') \delta \bm{\mathcal{K}}^{(n+1)}(t') , \label{eq:Picard2_Brownian}
\end{align}
and it is clear that they correspond to the iteration scheme of the preceding subsection.
Then $\mat{S}^{(n)}(t), - \delta \bm{\mathcal{K}}^{(n)}(t), -\delta \mat{K}^{(n)}(t)$ are completely monotone. By construction $\mat{S}^{(n)}(t=0) = \mat{S}$, $-\delta \bm{\mathcal{K}}^{(n)}(t=0) = \bm{\mathcal{D}} \bm{\mathcal{M}}(t=0) \bm{\mathcal{D}}$ etc. and
one infers  $\norm{ \mat{S}^{(n)}(t) } \leq \norm{ \bm{S}} , \norm{\delta \bm{\mathcal{K}}^{(n)}(t) } \leq \norm{ \bm{\mathcal{D}} }^2 \, \norm{ \bm{\mathcal{M}} } := K_1,
 \norm{ \delta \mat{K}^{(n)}(t)}  \leq K_1 L_1  $.
 Then the following bounds follow for $0\leq t \leq T < \infty $
\begin{align}\label{eq:S_estimate}
 \lefteqn{ \norm{ \mat{S}^{(n+1)}(t) - \mat{S}^{(n)}(t) } \leq \nonumber } \\
\leq & \norm{ \mat{D} } \, \norm{ \mat{S}^{-1} } \, \int_0^t \diff t'      \norm{ \mat{S}^{(n+1)}(t') - \mat{S}^{(n)}(t') } \nonumber \\
&  +
 T \norm{\mat{S}^{-1} } \, \norm{ \mat{S} }  \int_0^t \diff t' \norm{ \delta \mat{K}^{(n)}(t') - \delta \mat{K}^{(n-1)}(t') } \nonumber \\
&+ T K_1 L_1 \norm{\mat{S}^{-1}} \int_0^t \diff t' \norm{ \mat{S}^{(n+1)}(t') - \mat{S}^{(n)}(t') }  ,
\end{align}
\begin{align}\label{eq:K_estimate}
\lefteqn{ \norm{\delta \bm{\mathcal{K}}^{(n+1)}(t) - \delta \bm{\mathcal{K}}^{(n)}(t) } \leq \nonumber } \\
\leq & \norm{ \bm{\mathcal{D}} }^2 L_2   \norm{  \mat{S}^{(n)}(t) - \mat{S}^{(n-1)}(t) } \nonumber \\
& +  \norm{ \bm{\mathcal{D}}}\,   \norm{ \bm{\mathcal{M}} } \int_0^t \diff t'  \norm{  \delta \bm{\mathcal{K}}^{(n+1)}(t') - \delta \bm{\mathcal{K}}^{(n)}(t') }
\nonumber \\
& + \norm{  \bm{\mathcal{D}} }  K_1 L_2 \int_0^t \diff t'  \norm{  \mat{S}^{(n)}(t') - \mat{S}^{(n-1)}(t')} .
\end{align}
The presence of the  first term on the r.h.s of Eq.~\eqref{eq:K_estimate} requires to adapt the strategy to find suitable bounds. Upon substituting the bounds from Eq.~\eqref{eq:S_estimate}
the combination of  both equations yields estimates for $X^{(n)}(t) = \norm{ \mat{S}^{(n+1)}(t) - \mat{S}^{(n)}(t) } +  c\norm{ \bm{\mathcal{K}}^{(n+1)}(t) - \bm{\mathcal{K}}^{(n)}(t) }$ which now satisfies the inequalities
\begin{align}\label{eq:growth_control2}
 X^{(n)}(t) \leq L
\int_0^t \diff t' [ X^{(n)}(t') + X^{(n-1)}(t') + X^{(n-2)}(t') ],
\end{align}
for $n \geq 2$. Combining both equations yields estimates as in Eq.~\eqref{eq:growth_control} with the constant now replaced by $L =    (\norm{ \mat{D}} \,  \norm{ \mat{S}^{-1} }  + T L_1 \norm{  \mat{S} } \norm{  \mat{S}^{-1} }
 + T K_1 L_1 \norm{\mat{S}^{-1} } ) (1+ c \norm{ \bm{\mathcal{D}} }^2 L_2)  +c\norm{ \bm{\mathcal{D}} }\,  \norm{ \bm{\mathcal{M}} } + c \norm{ \bm{\mathcal{D}} } K_1 L_2 < \infty$.
The sequence $X^{(n)}(t)  $ is bounded by the constants $K = 2 \norm{\mat{S}} + 2 c K_1$ and can be controlled by
a suitable functional series $a^{(n)}(t)$ satisfying the inequalities
\begin{equation}\label{eq:a_sequence}
 a^{(n)}(t) \geq  L \int_0^t [ 2 a^{(n-1)}(t')+ a^{(n-2)}(t')] \diff t',
\end{equation}
  and $a^{(0)}(t) =  a^{(1)}(t) =  K$.  First,
 $X^{(0)}(t) \leq a^{(0)}(t)$,
$X^{(1)}(t) \leq a^{(0)}(t)$, $X^{(2)}(t) \leq a^{(0)}(t)$ and
 $X^{(1)}(t) \leq a^{(1)}(t)$, $X^{(2)}(t) \leq a^{(1)}(t)$ are obvious.
Then
$X^{(2)}(t) \leq L \int_0^t \diff t' [ 2 a^{(1)}(t')+ a^{(0)}(t')]  \leq  a^{(2)}(t)$.  The procedure can be continued to show that $X^{(n)}(t) \leq a^{(m)}(t)$ for $m\leq n$. It remains to find a suitable bound $a^{(n)}(t)$ satisfying Eq.~\eqref{eq:a_sequence} such that $\sum_n a^{(n)}(t) < \infty$. One can show by induction that $a^{(n)}(t) = K(3 \bar{L} \sqrt{t T})^{(n-1)}/\Gamma( (n+1)/2), n\in \mathbb{N}$ will do, provided the constant $\bar{L} \geq L +1/T$ is chosen large enough.  The remainder of the proof  follows the Newtonian case.

\section{Generalized Covariance Principle and the Maximum theorem}
In this section we demonstrate that the MCT equations  satisfy a covariance principle under a linear transformation of $\mat{S}(q,t)$. This covariance allows to prove the maximum theorem, which states that the long-time limit $\mat{S}(q,t\to \infty)$, provided it exists, can be calculated as maximal solution of a fixed-point equation. Solutions where these limits are identically zero are called ergodic and identified with fluid behavior, while non-ergodic ones correspond to (idealized) glass states. Then the maximum theorem allows calculating the phase diagram and in particular the glass-transition lines without solving the dynamic equations explicitly. Furthermore, since the fixed point equation is purely algebraic, the glass-transition lines originate as bifurcations associated with singular behavior in its vicinity.

For the case of a simple liquid these properties have been proven rigorously for Brownian dynamics~\cite{Goetze:1995}, and the proofs can be adapted also for the Newtonian case assuming the existence of a long-time limit of the ISF. For matrix-valued ISF and Brownian dynamics all properties can be properly generalized~\cite{Franosch:2002} relying on the notion of $C^*$ algebras. Here we shall show that for multiple decay channels the covariance principle can  be extended such that maximum theorem still holds, yet the transform for the memory kernels becomes non-linear and non-local in time.

\subsection{Fixed-point equation}
In the preceding section, we have shown that the solutions $\mat{S}(t)$ of the MCT equations are correlation functions, in particular they are positive-semidefinite for all times. Then the long-time limit, also referred to as glass-form factor,
\begin{align}
 \mat{F}(q) = \lim_{t\to \infty} \mat{S}(q,t) \succeq 0,
\end{align}
inherits this property, provided the limit exists. While this is guaranteed for Brownian dynamics (the solutions are non-negative, monotonically decreasing), in the Newtonian case all numerical solutions suggest that the limit also exist, yet no rigorous proof appears to be available. To proceed, we shall hence \emph{assume} that the limit exists also for Newtonian dynamics.

By the MCT functional, the memory kernel also displays a long-time limit
\begin{equation}\label{eq:memory_glass}
 \bm{\mathcal{N}}(q) := \lim_{t\to\infty} \bm{\mathcal{M}}(q,t) = \bm{\mathcal{F}}[\mat{F},\mat{F};q] \succeq 0,
\end{equation}
  which is again positive-semidefinite as it should be since microscopically the memory kernel is also a correlation function.
In the Laplace domain a non-vanishing long-time limit corresponds to a simple pole at zero frequency
$\hat{\bm{\mathcal{M}}}(z) = - z^{-1}\bm{\mathcal{N}}[1+ o(z^{0})]$ for $z\to 0$, where we again, here and in the following, suppress the dependence on the wave number. Then the representation of the current correlator both for Newtonian, Eq.~\eqref{eq:newtonian_dynamics}, as well as for Brownian dynamics, Eq.~\eqref{eq:Brownian_dynamics}, lead to the low-frequency behavior $\hat{\bm{\mathcal{K}}}(z) = z  \bm{\mathcal{N}}^{-1} + o(z)$. Thus the contraction
reads to leading order $\hat{\mat{K}}(z) = z \mat{N}^{-1} + o(z)$ with
\begin{align}\label{eq:effective_functional}
 [\mat{N}^{-1}]_{\mu\nu} = \sum_{\alpha \beta} q^{\alpha}_\mu [\bm{\mathcal{N}}^{-1}]^{\alpha \beta}_{\mu\nu} q^\beta_\nu \succeq 0.
\end{align}
Considered as a functional of the long-time limits $\mat{N}[\mat{F}]$ displays the properties of an \emph{effective} static mode-coupling functional in the space of matrices with mode indices $\mu,\nu$, as shown in Ref.~\cite{Lang:2012}. In particular, it preserves ordering $\mat{N}[\mat{E}] \succeq \mat{N}[\mat{F}]$ if $\mat{E} \succeq \mat{F}$.

The first equation of motion, Eq.~\eqref{eq:first_zwanzig}, shows that $\hat{\mat{S}}(z) = - z^{-1}\mat{F} [ 1+ o(z^{0})]$ for $z\to 0$, where the glass-form factor now satisfies~\cite{Lang:2012}
\begin{align}\label{eq:fixpoint}
 \mat{F} =& [\mat{S}^{-1} + \mat{S}^{-1} \mat{N}[\mat{F}]^{-1} \mat{S}^{-1} ]^{-1} \nonumber \\
=& \mat{S} - [\mat{S}^{-1} + \mat{N}[\mat{F}] ]^{-1}.
\end{align}
The Eqs.~\eqref{eq:memory_glass},\eqref{eq:effective_functional},\eqref{eq:fixpoint} constitute a closed set of equations called fixed-point equations and the long-time limit has to represent one of the solutions of these equations. The coupling of different wave numbers originates now from the effective static mode-coupling functional.
In general, the fixed-point equations allow for many solutions, in particular zero is always  a solution. Hence a criterion which of the solutions represents the long-time limit of the dynamic MCT equations is needed. For the case of single-decay channels, the answer is provided by the maximum theorem stating that the long-time limit is represented by the maximal solution~\cite{Goetze:Complex_Dynamics,Goetze:1999,Franosch:2002}.
More precisely, out of all solutions $\bar{\bar{\mat{F}}} \succeq 0$ that are positive-semidefinite  there is a unique solution $\bar{\mat{F}}$
that is larger than any of the other  $\bar{\bar{\mat{F}}}$,  $\bar{\mat{F}} \succ \bar{\bar{\mat{F}}}$ and this is the one  corresponding to the glass-form factor. The maximal solution can be found by a monotone iteration scheme.

For multiple decay channels, it has been shown~\cite{Lang:2012} that the fixed-point equations admit for a maximal solution, which again can be determined by an iteration scheme. Here we prove that it also  corresponds to the long-time limit of the dynamic MCT equations.

\subsection{Effective mode-coupling theory functional and generalized covariance principle}
The strategy consists of properly generalizing the covariance principle demonstrated for one-component systems~\cite{Goetze:1995} and mixtures~\cite{Franosch:2002}. Therefore we shall construct an effective dynamic mode-coupling functional $\mat{M}(t)$ acting only in the space of matrices with mode indices. The effective dynamic functional will be shown to be a proper extension of the effective static  functional used in the preceding subsection to preserve the properties of correlation functions or completely monotone functions. This effective dynamic  functional is then the suitable starting point  to apply the covariance idea.

First, we consider \emph{Newtonian} dynamics and define the kernel $\hat{\mat{M}}(z)$ in the Laplace domain via
\begin{align}\label{eq:effective_kernel}
 \left[z \mat{J}^{-1}  + \hat{\mat{M}}(z) \right]^{-1} := - \hat{\mat{K}}(z).
\end{align}
 In order to keep the argument simple, the regular damping has been discarded $\bm{\mathcal{D}}^{-1} \equiv 0$. In Appendix~\ref{appendix:representation} it is shown that $\hat{\mat{M}}(z)$ is well defined and again corresponds to a correlation function.
The current kernel $\hat{\mat{K}}(z)$  itself is the  contraction of $\hat{\bm{\mathcal{K}}}(z)$ which is obtained from  the memory kernels $\hat{\bm{\mathcal{M}}}(z)$ by Eq.~\eqref{eq:newtonian_dynamics}. Then $\hat{\mat{M}}(z)$ can be viewed itself as a functional of the intermediate scattering functions and we refer to it as the effective dynamic  MCT functional. Specializing Eq.~\eqref{eq:effective_kernel} to the limit $z\to 0$ and going through the arguments of the preceding subsection again shows that the effective dynamic functional reduces to the effective static functional $\mat{N}$.  Casting
Eq.~\eqref{eq:effective_kernel} in the temporal domain
\begin{equation}
  \int_0^t \mat{J} \mat{M}(t') \mat{K}(t-t')\diff t' = -\partial_t \mat{K}(t) ,
\end{equation}
shows that given $\mat{K}(t')$ for times $t'\leq t$ the effective MCT functional can be determined up to time $t$ by solving a standard (matrix-valued) Volterra integral equation of the first kind. Thus the effective dynamic MCT functional is a \emph{causal} smooth functional of the intermediate scattering functions.
Whereas the original memory kernel is local in time this is in general no longer the case for the effective MCT functional originating from multiple relaxation kernels as can be checked by a high-frequency expansion. To emphasize the dependence on the time-dependent past of the intermediate scattering function,  we indicate the effective mode-coupling functional by $\mat{M}[\{ \bm{S}\}](t)$.
However its  initial value fulfills  that $\mat{J}^{-1} \mat{M}(t=0) \mat{J}^{-1}$  is merely the contraction of
$\bm{\mathcal{J}}^{-1} \bm{\mathcal{M}}(t=0) \bm{\mathcal{J}}^{-1}$, as follows from the short-time expansion.

The next step is to reformulate the iteration scheme of the preceding section  in terms of the effective memory kernel. A simple calculation shows that the
map
\begin{align}\label{eq:mapping}
&\bm{\mathcal{H}} : \left(\hat{\mat{S}}(z), \hat{\mat{M}}(z)\right) \mapsto \nonumber \\
&   \left(-\mat{S}/z + [ z \mat{S}^{-1} - z^3 \mat{J}^{-1} - z^2 \hat{\mat{M}}(z)]^{-1}, \hat{\mat{M}}[\{\mat{S}\}](z) \right)   ,
\end{align}
generates the same sequence of approximants as  above provided it is initialized with $(\hat{\mat{S}}^{(0)}(z) = -\mat{S}/z, \hat{\mat{M}}^{(0)}(z)=0)$.
In particular, the sequence was shown never to leave the space of correlation functions and
to converge to the fixed point solution of $\bm{\mathcal{H}}(\hat{\mat{S}}(z), \hat{\mat{M}}(z))=(\hat{\mat{S}}(z), \hat{\mat{M}}(z))$ which is the unique solution of the MCT equations.

Assume now that $\bar{\bar{\mat{F}}}$ with $\mat{S} \succeq\bar{\bar{\mat{F}}}\succeq 0$ is a solution of the fixed point equation, Eqs.~\eqref{eq:memory_glass},\eqref{eq:effective_functional},\eqref{eq:fixpoint},
then we define a mapping
\begin{align}\label{eq:tilde}
 \tilde{}: \Big(\mat{S}(t), \mat{M}[\{ \mat{S} \} ](t) \Big) \mapsto
& \Big(\tilde{\mat{S}}(t) = \mat{S}(t)-\bar{\bar{\mat{F}}}, \nonumber \\
& \tilde{\mat{M}}[\{ \tilde{\mat{S}} \} ](t) = \mat{M}[ \{ \mat{S} \}](t) - \mat{N}[\bar{\bar{\mat{F}}}]\Big).
\end{align}

To make progress we make the assumption that   long-time limits, $\mat{F} = \lim_{t\to \infty} \mat{S}(t), \lim_{t\to \infty} \mat{M}(t)$
of  the correlation functions $\mat{S}(t), \mat{M}(t)$
emerging as fixed points of the map, Eq.~\eqref{eq:mapping}, exist. Analyzing the low-frequency behavior of Eq.~\eqref{eq:effective_kernel} and employing the definition in Eq.~\eqref{eq:effective_functional}
reveals that $\lim_{t\to\infty} \mat{M}(t) = \mat{N}[\mat{F}]$.
The spectral representation, Eq.~\eqref{eq:spectral_representation},  further shows that they correspond to  a jump in the spectral measure  at zero frequency, $\lim_{t\to \infty} {\mat{S}}(t) = {\mat{R}}(\{ 0 \}) \succeq 0$, and is positive-semidefinite and correspondingly for $\mat{M}(t)$. Considering $\tilde{\mat{M}}[\{ \tilde{\mat{S}} \}]$ as a functional of  $\tilde{\mat{S}}(t) $
we have to show that it maps correlation functions $\tilde{\mat{S}}(t)$ to correlation functions $\tilde{\mat{M}}(t)$. Since the spectral measures $\mat{R}_{\mat{M}}, \mat{R}_{\tilde{\mat{M}}}$ associated to $\mat{M}(t)$ and $\tilde{\mat{M}}(t)$ differ only by a jump at zero frequency $\mat{R}_{\tilde{\mat{M}}}(\{  0 \} ) = \mat{R}_{\mat{M}}(\{ 0 \}) - \mat{N}[\bar{\bar{\mat{F}}}]$, what remains to  be shown is that $\lim_{t\to\infty} \tilde{\mat{M}}(t) = \lim_{t\to\infty} \mat{M}(t) - \mat{N}[\bar{\bar{\mat{F}}}] = \mat{N}[\mat{F}]- \mat{N}[\bar{\bar{\mat{F}}}] \succeq 0$.  However by assumption $\tilde{\mat{S}}(t)$ is also a correlation function, thus its long-time limit fulfills $\lim_{t\to\infty} \tilde{\mat{S}}(t) = \mat{F}-\bar{\bar{\mat{F}}} \succeq 0$.  Then the desired property follows since the functional $\mat{N}[\mat{F}]$ preserves ordering.

The next step is to construct a mapping $\tilde{\bm{\mathcal{H}}}$ such that the diagram
\begin{align}\label{eq:Diagram}
\begin{xy}
  \xymatrix{
     ( \hat{\mat{S}}(z) ,  \hat{\mat{M}}(z) )
\ar[r]^{\bm{\mathcal{H}}} \ar[d]_{\tilde{}}    &  \bm{\mathcal{H}}[\hat{\mat{S}}(z),\hat{\mat{M}}(z)] \ar[d]^{\tilde{}}  \\
      (\hat{\tilde{\mat{S}}}(z),\hat{\tilde{\mat{M}}}(z)) \ar[r]^{\tilde{\bm{\mathcal{H}}}}             &    \tilde{\bm{\mathcal{H}}}[\hat{\tilde{\mat{S}}}(z),\hat{\tilde{\mat{M}}}(z)] ,
  }
\end{xy}
\end{align}
commutes. Explicit calculation shows that this is fulfilled with
\begin{align}
&\tilde{\bm{\mathcal{H}}} : \left(\hat{\tilde{\mat{S}}}(z), \hat{\tilde{\mat{M}}}(z)\right) \mapsto \nonumber \\
&   \left(-\tilde{\mat{S}}/z + [ z \tilde{\mat{S}}^{-1} - z^3 \mat{J}^{-1} - z^2 \hat{\tilde{\mat{M}}}(z)]^{-1}, \hat{\tilde{\mat{M}}}[\{\tilde{\mat{S}}\}](z) \right)   ,
\end{align}
and $\tilde{\mat{S}} = \mat{S} - \bar{\bar{\mat{F}}}$ as follows from Eq.~\eqref{eq:tilde}. The preceding equations show that the MCT equations
 are \emph{covariant} with respect to the linear transform $\tilde{}$, i.e. all equations assume the same form and aquire merely a $\tilde{}$. Note that the current matrix $\mat{J}$
remains the same, which becomes obvious upon inspection of  the short-time expansion, Eq.~\eqref{eq:short_time_Newton}.
Comparison with Ref.~\cite{Lang:2012} shows that this mapping is the proper generalization of the renormalized mode-coupling functional to the case of finite frequencies $z$.

For \emph{Brownian} motion the argument runs essentially along the same path. We define the effective memory kernel $\hat{\mat{M}}(z)$ in the Laplace domain for the Brownian case via
\begin{align}\label{eq:effective_kernel_brown}
 \left[i \mat{D}^{-1}  + \hat{\mat{M}}(z) \right]^{-1} :=  -\hat{\mat{K}}(z) = - i \mat{D} -\delta\hat{\mat{K}}(z).
\end{align}
It is shown in Appendix~\ref{appendix:representation_Brown} that
 $\hat{\mat{M}}(z)$ is well-defined and corresponds to a completely monotone function. Again the effective memory kernel should be viewed as a non-local causal smooth functional of the intermediate scattering functions, $\mat{M}[\{\mat{S} \} ](t)$.
 However, its  initial value  fulfills that  $\mat{D}^{-1} \mat{M}(t=0) \mat{D}^{-1}$   is merely the contraction of
$\bm{\mathcal{D}}^{-1} \bm{\mathcal{M}}(t=0) \bm{\mathcal{D}}^{-1}$.
We also note that the definitions of the effective kernel for the Newtonian and Brownian case differ from each other, nevertheless considered as functional of $\mat{S}(t)$ they are expected to display the same long-time behavior.  By construction they both reduce to the  effective static functional in the long-time limit.

The mapping $\bm{\mathcal{H}}$ differs only by notation from the Newtonian case
\begin{align}
&\bm{\mathcal{H}} : \left(\hat{\mat{S}}(z), \hat{\mat{M}}(z)\right) \mapsto \nonumber \\
&   \left(-\mat{S}/z + [ z \mat{S}^{-1} - z^2 i\mat{D}^{-1} - z^2 \hat{\mat{M}}(z)]^{-1}, \hat{\mat{M}}[\{\mat{S}\}](z)  \right)   .
\end{align}

A fixed-point solution $\bar{\bar{\mat{F}}}$ induces the mapping ${}^{\tilde{}}$, as in Eq.~\eqref{eq:tilde}.
Provided  $\tilde{\mat{S}}(t)$ is again completely monotone, then $\tilde{\mat{M}}(t)$ is also completely monotone  as can be seen as follows. By the representation theorem
 $\mat{M}(t)-\lim_{t\to \infty}\mat{M}(t)$  is  completely monotone (note that the long-time limit is guaranteed to exist), thus $\tilde{\mat{M}}(t)$ is completely monotone if $\lim_{t\to\infty}\mat{M}(t) - \mat{N}[\bar{\bar{\mat{F}}}] \succeq 0$. The remainder of the argument is as in the Newtonian case.

The covariance of the MCT equation expressed in terms of a commuting diagram, Eq.~\eqref{eq:Diagram} is achieved by
\begin{align}
&\tilde{\bm{\mathcal{H}}} : \left(\hat{\tilde{\mat{S}}}(z), \hat{\tilde{\mat{M}}}(z)\right) \mapsto \nonumber \\
&   \left(-\tilde{\mat{S}}/z + [ z \tilde{\mat{S}}^{-1} - z^2 i \mat{D}^{-1} - z^2 \hat{\tilde{\mat{M}}}(z)]^{-1}, \hat{\tilde{\mat{M}}}[\{\tilde{\mat{S}}\}](z)  \right)   ,
\end{align}
and again $\tilde{\mat{S}} = \mat{S} - \bar{\bar{\mat{F}}}$.

\subsection{Maximum principle}
The covariance property shows that the problem to find a solution $\mat{S}(t)$ of the MCT equations is the same as finding a solution $\tilde{\mat{S}}(t) =\mat{S}(t)-\bar{\bar{\mat{F}}}$ for the mapped problem, provided $\bar{\bar{\mat{F}}}\succeq 0$ is a solution of the fixed-point equations. Remember that for the Newtonian case we had to use the additional assumption that the long-time limit exists. But the existence and uniqueness of such a solution has been shown by the iteration scheme in the temporal domain. Hence we conclude that $\tilde{\mat{S}}(t)$ is uniquely determined and corresponds to a  correlation function (where the long-time limit exists by assumption) in the case of Newtonian dynamics and to a completely monotone function for Brownian dynamics.

For the long-time limits $\mat{F} = \lim_{t\to\infty}\mat{S}(t)\succeq 0$, $\tilde{\mat{F}} = \lim_{t\to\infty}\tilde{\mat{S}}(t)\succeq 0$ this implies the conclusion $\mat{F} \succeq \bar{\bar{\mat{F}}}$. Hence the long-time limits are not smaller than any solution of the fixed-point equations. Yet the long-time limits are a solution of the fixed-point equations themselves and therefore correspond to the maximal solution $\mat{F} \equiv \bar{\mat{F}}$.
The observation is referred to as maximum property and allows to determine the long-time limit of the MCT equations by solving algebraic equations rather than coupled integro-differential equations.

By using the effective MCT functional we have reduced the problem of discussing the mathematical properties of the solutions of the MCT to the case of a single decay channel. The corresponding results can be taken over directly. In particular, consider the linearization $\mat{\Psi}[\delta \mat{F}]$ as obtained from
\begin{equation}
[\mat{S}^{-1}+ \mat{N}[\mat{F}]]^{-1} - [ \mat{S}^{-1}+ \mat{N}[\mat{F} +\delta \mat{F}] ]^{-1} =  \mat{\Psi}[\delta \mat{F}] + \mathcal{O}(\delta \mat{F})^2,
\end{equation}
where $\mat{F} \succeq 0$ is the maximal solution of the fixed-point equation. One readily
shows that $\mat{\Psi}: \mathcal{A}_0^M \to \mathcal{A}_0^M$ is a linear positive map on a C* algebra, i.e. it fulfills $\mat{\Psi}[\delta \mat{F}] \succeq 0$ if $\delta \mat{F} \succeq 0$. In Ref.~\cite{Franosch:2002} it was shown by a proper extension of the Frobenius-Perron theorem that $\mat{\Psi}$ has a maximum non-degenerate eigenvalue not exceeding unity in the generic case  that $\mat{\Psi}$ is irreducible. The manifold in the phase diagram where this eigenvalue becomes unity is identified with the glass-transition lines and the associated bifurcation behavior has to be of the $A_\ell$ type according to the classification of Arnol'd~\cite{Arnold:1975}. The generic case corresponds to the fold bifurcation scenario $A_2$,
and explicit expressions  for the exponent parameter have been worked out~\cite{Voigtmann:thesis,Hajnal:2009}. For the case of mixtures   asymptotic solutions with scaling properties hold in the vicinity of the glass-transition lines similar to the one-component case~\cite{Voigtmann:thesis,Weysser:2011}.

\section{Summary and Conclusions}

In general a density mode can decay into more than one channel such that the associated particle current density  naturally splits into different parts. The representation in terms of memory kernels then suggests to consider parallel relaxation which implies a mathematically different structure of the equations of motion. Within the mode-coupling approach the force kernels are again approximated by positive superpositions of products of density correlation functions at the same instant of time. Our work provides proofs that the  solutions of this generalized MCT still  respect the constraints of probability theory and purely relaxational dynamics, extending the ideas introduced in the one-component case with a single decay channel~\cite{Haussmann:1990,Goetze:1995,Goetze:Complex_Dynamics} and their generalization to matrix-valued correlation functions~\cite{Franosch:2002}.

The proof of existence and uniqueness of solutions  relies on an iteration scheme similar to ordinary differential equations and the generalization to multiple decay channels requires only  mild adaptions. Similarly, the ideas to show that the iteration does not leave the space of correlation functions or completely monotone functions are directly transferable from the matrix case. The shown properties underline that the MCT approach is robust and encodes a series of natural requirements on any theory of dynamical phenomena without fine tuning.

Much stronger conclusions follow from a property observed by G\"otze  known as covariance principle~\cite{Goetze:LesHouches,Goetze:Complex_Dynamics}. First a fixed-point equation is derived whose solutions are candidates for the long-time limits, i.e. glass form factors, then it is shown that the equations of motion determining the dynamics 'on top' of this candidate is of the same form as the original problem. Since the solutions of the transformed problem are again correlation functions or completely monotone, one can define a semi-ordering for glass-form factors such that the long-time limits of the dynamic equations coincide with the maximal solution of the fixed-point equations.

While the matrix-valued theory displays this covariance property in  a straightforward generalization of the single-component theory, this is no longer the case for multiple decay channels. In fact, one can convince oneself that application of a linear transform to the density correlation function and the MCT kernels does not lead to form-invariant equations of motion. The key idea has been to introduce an effective (single-decay channel) MCT functional to map the theory to a case where the covariance principle is known to apply. On the level of the glass-form factors this was implemented in Ref.~\cite{Lang:2012} and it was shown that the effective MCT functional preserves ordering. The extension to the dynamics comes at the price that the functional is no longer local in time, yet it remains causal in the sense that  to determine the effective kernel at a given time
 knowledge of the density correlation functions at earlier times is sufficient.
With this prerequisite, we have shown a generalized covariance principle, such that essentially all of the consequences elaborated
for the one-component case can be taken over. In particular, the maximum property follows, stating that the long-time limits of the dynamic theory can be calculated without solving the dynamic equations explicitly. Rather it is sufficient to compute the maximal solution of the fixed-point equation. This in turn permits to distinguish ergodic liquid-like solutions from non-ergodic idealized states and thus to construct a non-equilibrium state diagram. The glass-transition lines are identified with bifurcations of  the fixed-point equation and they have to be of $A_\ell$ type in the classification by Arnol'd~\cite{Arnold:1975}, the simplest being the $A_2$ fold bifurcation.

We have not pursued to reproduce properties such as a non-zero radius of convergence for a short-time expansion, the existence of a power series for small frequencies for control parameter off the glass-transition lines, or the emergence of scaling laws in the vicinity of the glass-transition. Since the effective mode-coupling functional is non-local in time, the proofs presented for the one- and multicomponent liquid~\cite{Goetze:1995,Franosch:2002,Goetze:Complex_Dynamics,Franosch:1997} do not readily apply, however it appears promising to repeat the essential steps on the level of the multiple channel description.

An additional term linear in the intermediate scattering function may be added in the mode-coupling functional, as it occurs naturally for the interaction of liquid particles with a frozen disorder, see for example ~\cite{Krakoviack:2005,Krakoviack:2007,Schnyder:2011, Spanner:2013}. Provided, the full functional still satisfies the constraints formulated in Sec.~\ref{sec:functional}, all conclusions for the general properties of the MCT-solutions for the multichannel MCT remain unaffected. More generally, all non-linear functionals of the intermediate scattering function which map correlation functions in the case of Newtonian dynamics, or purely relaxational functions for Brownian dynamics, onto functions with the same properties, can be included without modification.
Similarly, the extension of the proofs to the incoherent dynamics coupled to the collective motion should not require new ideas. 

The MCT-equations may be equipped by an additional a priori known regular damping term, see Eq.~\eqref{eq:MCT_M}, which captures fast processes that are not included within the MCT-functional and are smoothly embedded into our analysis. Generically, intriguing glass transition scenarios may occur, if one of the channels dominates the caging mechanism. For example partial freezing only of the translational degrees of freedom for dipolar hard spheres~\cite{Scheidsteger:1997} or for the collective rotational degrees of freedom for elongated ellipsoids~\cite{Letz:2000} have been uncovered.

The MCT equations for multiple decay channels may challenge the replica theory for structural glasses~\cite{Mezard:1996}, which is a purely static theory. Besides predicting a thermodynamic glass transition of Kauzmann type it also allows to deduce a singularity for a dynamical transition by deriving an equation for the glass-form factors. For a hard-sphere liquid in high dimensions  $d \rightarrow \infty$ it has recently been shown that the critical volume fraction $\varphi^{\rm replica}_{\rm dyn} (d)$ from replica theory~\cite{Parisi:2006} and $\varphi^{\rm MCT} (d)$ from MCT~\cite{Miyazaki:2010,Schmid:2010} do not coincide. Using a different closure for the calculation of the static input into replica theory Szamel~\cite{Szamel:2010b} has demonstrated that replica theory yields the identical equation for the glass-form factors as MCT does. However, since replica theory lacks a dynamical origin, it seems impossible to prove that the long-time limit of the intermediate scattering functions equals the
maximum
solution for the glass-form factors. In
addition since replica theory only considers density and no current density, probably, its form does not involve a channel index. Consequently, its fixed-point equation for the glass-form factors  will not depend on the number of decay channels, in strong contrast to the corresponding equation from MCT. How replica theory and MCT in case of more than one decay channel could be reconciled is far from being  obvious.

\appendix
\section{Representation Theorem for Newtonian Dynamics}\label{appendix:representation}
In this appendix we show that a matrix-valued correlation function can be represented in the Laplace domain via another correlation function.

\paragraph*{Theorem:} If $\hat{\mat{K}}(z)$ is the Fourier-Laplace transform of a matrix-valued correlation function with asymptotic expansion
\begin{equation}\label{eq:asymptotic_expansion}
\hat{\mat{K}}(z) = - z^{-1} \mat{J} + \mathcal{O}(z^{-3}), \qquad z\to \infty,
\end{equation}
with $\mat{J} \succ 0$, then there exists a representation
\begin{equation}
\hat{\mat{K}}(z) = - [z \mat{J}^{-1} + \hat{\mat{M}}(z) ]^{-1},
\end{equation}
such that $\hat{\mat{M}}(z)$ corresponds again to a correlation function.

\paragraph*{Proof:} The proof is a generalization of the corresponding property in the scalar case (see Akhiezer~\cite{Akhiezer:Classical_Moment_Problem}, p.111, Lemma 3.3.6).

By the spectral representation theorem,
\begin{equation}
\hat{\mat{K}}(z) = \int (\Omega-z)^{-1}  \mat{R}_{\mat{K}}(\diff\Omega),
\end{equation}
with $\mat{R}_{\mat{K}}(\Omega)$ the associated self-adjoint matrix-valued measure. From the asymptotic expansion, Eq.~\eqref{eq:asymptotic_expansion}, one infers $\mat{J} =
\int  \mat{R}_{\mat{K}}(\diff \Omega)$ and by assumption $\mat{J}$ is positive-definite. Next, solve formally for $\hat{\mat{M}}(z)$
\begin{equation}\label{eq:solveM}
\hat{\mat{M}}(z) = -z \mat{J}^{-1} - \hat{\mat{K}}(z)^{-1}, \qquad z\in \mathbb{C}_+.
\end{equation}
In order for this being well-defined, $\hat{\mat{K}}(z)$ has to be invertible in the complex upper half plane. Assume for the moment the contrary. Then, there is a vector $\bm{y}$ not identically to zero and for a  $z_1\in \mathbb{C}_+$ with $\hat{\mat{K}}(z_1) \bm{y} = 0$, implying that the complex analytic function $\bm{y}^\dagger \hat{\mat{K}}(z) \bm{y}$ displays a zero in the upper half plane. However, the representation
\begin{align}\label{eq:representation_Imag}
 \Imag[\hat{\mat{K}}(z)]  = \int \frac{\Imag[z]  \mat{R}_{\mat{K}}  (\diff \Omega)}{|\Omega-z|^2} \succeq 0
\end{align}
reveals that $ \Imag[\bm{y}^\dagger \hat{\mat{K}}(z) \bm{y}]\geq 0$ is non-negative and harmonic and by the mean-value property a zero can occur only if  $\bm{y}^\dagger \hat{\mat{K}}(z) \bm{y} \equiv 0$ for all $z\in\mathbb{C}_+$. The latter case is excluded by the assumption on the asymptotic expansion, Eq. \eqref{eq:asymptotic_expansion}.

Thus $\hat{\mat{M}}(z)$ is well-defined in the upper half plane. To demonstrate that it corresponds to a correlation function, we show again properties (1)-(4) of Sec.~\ref{Sec:correlation_function}.
The explicit expression, Eq.~\eqref{eq:solveM} then shows that $\hat{\mat{M}}(z)$ is analytic in $\mathbb{C}_+$, proving property (1).
Property (2) follows immediately from Eq.~\eqref{eq:solveM} and $\hat{\mat{K}}(-z^*) = - \hat{\mat{K}}(z)^\dagger$, since $\mat{K}(t)$ is a correlation function.
From the asymptotic expansion, Eq.~\eqref{eq:asymptotic_expansion}, one finds
\begin{align}\label{eq:Mestimate}
\hat{\mat{M}}(z) =& - z \mat{J}^{-1} - [-z^{-1} \mat{J} + \mathcal{O}(z^{-3})]^{-1} \nonumber \\
=& -z \mat{J}^{-1} + z \mat{J}^{-1} [1 + \mathcal{O}(z^{-2})]^{-1} = \mathcal{O}(z^{-1}),
\end{align}
which implies property (3). From, Eq.~\eqref{eq:solveM}, the imaginary part can be represented as
\begin{equation}
\Imag[\hat{\mat{M}}(z)] = - \mat{J}^{-1} \Imag[z] + \hat{\mat{K}}^\dagger(z)^{-1} \Imag[\hat{\mat{K}}(z)] \hat{\mat{K}}(z)^{-1}.
\end{equation}
The right-hand side is positive-semidefinite as can be seen as follows. To any vector $\bm{y}$ define the vector
\begin{equation}
\bm{x}(\Omega) = (\Omega-z)^{-1} \hat{\mat{K}}(z)^{-1} \bm{y} - \mat{J}^{-1} \bm{y}.
\end{equation}
Then
$\int \bm{x}(\Omega)^\dagger  \mat{R}_{\mat{K}}(\diff \Omega) \bm{x}(\Omega) \geq 0$.
Upon expanding, one finds
\begin{align}
\bm{y}^\dagger \hat{\mat{K}}^\dagger(z)^{-1} \int \frac{ \mat{R}_{\mat{K}}(\diff \Omega)}{|\Omega-z|^2} \hat{\mat{K}}(z)^{-1} \bm{y} - \bm{y}^\dagger \mat{J}^{-1} \bm{y} \geq 0.
\end{align}
Last, the representation of  $\Imag[\hat{\mat{K}}(z)] $, Eq.~\eqref{eq:representation_Imag}, shows that
the kernel $\hat{\mat{M}}(z)$ displays positive-semidefinite imaginary part, and
property (4) also holds. $\square$

\section{Representation theorem for Brownian dynamics}\label{appendix:representation_Brown}
In this appendix we show that the effective memory kernel
\begin{equation}
\hat{\mat{M}}(z) =-\hat{\mat{K}}(z)^{-1} - i \mat{D}^{-1},
\end{equation}
is well-defined for frequencies $z\in \mathbb{C}\setminus i \mathbb{R}^{-}$ and is
completely monotone provided $-\delta\hat{\mat{K}}(z)= i \mat{D}-\hat{\mat{K}}(z)$ shares the same property and $ \Imag[ \hat{\mat{K}}(z)] \succ 0 $ for $z\in \mathbb{C}_+$.

\paragraph*{Proof:}
 If $\delta \hat{\mat{K}}(z) \equiv 0$ then $\hat{\mat{M}}(z) \equiv 0$ and nothing needs to be shown.

First, by assumption $\Imag[\hat{\mat{K}}(z) ] \succ 0$ for $z\in \mathbb{C}_+$ and $\hat{\mat{K}}(z)$ is invertible in the complex upper half plane. Next,
we show that $\hat{\mat{K}}(z) = i\mat{D}+ \delta\hat{\mat{K}}(z)$ is invertible also for $z\in \mathbb{C}\setminus i \mathbb{R}^-$.
Assume for the moment the contrary. Then there is a non-zero vector $\bm{y}$ with $[ i \mat{D} + \delta \hat{\mat{K}}(z)] \bm{y} =0$.
But since $-\delta \hat{\mat{K}}(z)$ corresponds to a completely monotone function, there is a self-adjoint complex measure $\mat{a}(\gamma)$ with
\begin{equation}
-\delta\hat{\mat{K}}(z) = \int_0^{\infty} \frac{-1}{z+i\gamma}  \mat{a}(\diff \gamma),
\end{equation}
and one finds
\begin{align}
& \Real[ \bm{y}^\dagger (i \mat{D} + \delta\hat{\mat{K}}(z) ) \bm{y} ] =
\Real[ \bm{y}^\dagger  \delta\hat{\mat{K}}(z)  \bm{y} ] \nonumber \\
=& \int_{0}^\infty \frac{\Real[z]}{|z+ i \gamma|^2} \bm{y}^\dagger  \mat{a}(\diff \gamma) \bm{y} \neq 0 \qquad \text{for } \Real[z] \neq 0.
\end{align}
which shows that $\hat{\mat{K}}(z)^{-1}$ is well-defined now for all $z\in \mathbb{C}\setminus i \mathbb{R}^-$.

The analytic properties (1*), (2*)  are inherited from the corresponding ones for $\delta\hat{\mat{K}}(z)$. Since $\lim_{z\to \infty} \hat{\mat{K}}(z) = i \mat{D}$ one finds $\lim_{z\to\infty} \hat{\mat{M}}(z) = 0$ which implies (3*). Last, (4*) follows from
\begin{equation}
 \Real[\hat{\mat{M}}(z)]  = \hat{\mat{K}}^\dagger(z)^{-1} \Real[-\delta\hat{\mat{K}}(z)] \hat{\mat{K}}(z)^{-1} \succeq 0  ,
\end{equation}
for $\Real[z]<0$ since $-\delta \hat{\mat{K}}(z)$ corresponds to a completely monotone function. By the representation theorem properties (1*)-(4*) for  $\hat{\mat{M}}(z)$ imply that it corresponds to the Fourier-Laplace transform of a completely monotone (matrix-valued) function. $\square$
\begin{acknowledgments}
It is a great pleasure to thank Wolfgang G\"otze for numerous discussions highlighting the implications of a different mathematical structure of multiple decay channels, in particular, the violation of the covariance principle in the strict sense.
This work has been supported by the
Deutsche Forschungsgemeinschaft DFG via the  Research Unit FOR1394 ``Nonlinear Response to
Probe Vitrification''.
\end{acknowledgments}

\bibliographystyle{apsrev4-1}



\end{document}